\documentclass[prd,aps,twocolumn,amsmath,nofootinbib,superscriptaddress]{revtex4-2}

\usepackage{amsmath}
\usepackage{amsfonts}
\usepackage{amssymb}	
\usepackage{graphicx}
\usepackage{bm}
\usepackage{color}
\usepackage{commath}
\usepackage{stmaryrd}
\allowdisplaybreaks

\usepackage{hyperref}
\hypersetup{
  colorlinks=true,        
  linkcolor=blue,         
  citecolor=cyan,         
  urlcolor=cyan           
}
\usepackage{times}

\newcommand{\ie}{i.e.,~}
\newcommand{\eg}{e.g.,~}
\begin{document}

\title{Horizon-penetrating form of parametrized metrics for static and
stationary black holes}

\author{Yixuan Ma}
\affiliation{Institut f{\"u}r Theoretische Physik, Max-von-Laue-Strasse
1, 60438 Frankfurt, Germany}

\author{Luciano Rezzolla}
\affiliation{Institut f{\"u}r Theoretische Physik, Max-von-Laue-Strasse
1, 60438 Frankfurt, Germany}
\affiliation{Frankfurt Institute for Advanced Studies,
Ruth-Moufang-Strasse 1, 60438 Frankfurt, Germany}
\affiliation{School of Mathematics, Trinity College, Dublin 2, Ireland}

\date{\today}

\begin{abstract}
The Rezzolla-Zhidenko (RZ) and Konoplya-Rezzolla-Zhidenko (KRZ)
frameworks provide an efficient approach to characterize agnostically
spherically symmetric or stationary black-hole spacetimes in arbitrary
metric theories. In their original construction, these metrics were
defined only in the spacetime region outside of the event horizon, where
they can reproduce any black-hole metric with percent precision and
a few parameters only. At the same time, numerical simulations of
accreting black holes often require metric functions that are regular
across the horizon, so that the inner boundary of the computational
domain can be placed in a region that is causally disconnected from the
exterior. We present a novel formulation of the RZ/KRZ parametrized
metrics in coordinate systems that are regular at the horizon and defined
everywhere in the interior. We compare the horizon-penetrating form of
the KRZ and RZ metrics with the corresponding forms of the Kerr metric in
Kerr-Schild coordinates and of the Schwarzschild metric in
Eddington-Finkelstein coordinates, remarking the similarities and
differences. We expect the horizon-penetrating formulations of the RZ/KRZ
metrics to represent new tools to study via simulations the physical
processes that occur near the horizon of an arbitrary black hole. 
\end{abstract}

\maketitle

\section{Introduction}
\label{sec:introduction}

The last few years have provided compelling evidence that black holes as
predicted by Einstein's general relativity are perfectly compatible with
gravitational-wave~\cite{Abbott2016fw} and
electromagnetic~\cite{Akiyama2019_L1_etal, EHT_SgrA_PaperI_etal}
observations. Yet, because of the uncertainties accompanying these
observations, there is still large room for alternative interpretations
within other theories of gravity (see, \eg Refs.~\cite{Mizuno2018,
Kocherlakota2021, Abbott2021b, EHT_SgrA_PaperVI_etal, Chattearjee2023}).

Because of the wide variety of existing alternative theories of gravity,
and to avoid the impractical approach in which a validation of
observations is made on a case-by-case manner for every single theory,
model-independent representations of generic black-hole spacetime have
been proposed to measure the deviation from general relativity of
alternative theories of gravity. In this way, it is in principle possible
to invoke astronomical observations to constrain possible deviations
between different black-hole geometries~\cite{Vigeland2011}.  A first
attempt in this direction is the parametrization of rotating black holes
by Johannsen and Psaltis~\cite{Johannsen2011} and
Johannsen~\cite{Johannsen2013}, who expanded the deviation from Kerr
metric in terms of a Taylor series of the dimensionless compactness
parameter $M / r$, where $M$ is the black hole mass and $r$ is a generic
radial coordinate. Despite some of its expansion coefficients can be
observationally constrained, such an approach requires an infinite number
of parameters with equal importance, and is only able to reproduce small
deviations from general relativity~\cite{Cardoso2014}.

These shortcomings were addressed first for nonrotating black holes by
the Rezzolla-Zhidenko (RZ) parametrization~\cite{Rezzolla2014}, which
expresses the deviation of a generic spherically symmetric metric from
Schwarzschild metric in terms of a Pad{\'e} expansion of a compactified
radial coordinate\footnote{The general formulation of the parametrized
metric makes it applicable also to non-vacuum spacetimes, such as those
involving a compact star or a boson star~\cite{Kocherlakota2020}.}. The
superior convergence properties of the continued-fraction expansion
allows one to approximate arbitrary black holes in alternative theories
reaching a percent precision with only a few expansion
coefficients~\cite{Kocherlakota2020, Kocherlakota2022}. The extension of
this approach to stationary black-hole spacetimes was later obtained with
the Konoplya-Rezzolla-Zhidenko (KRZ)
parametrization~\cite{Konoplya2016a}. The KRZ metric adopts the same
continued fraction expansion in the radial direction, and a Taylor
expansion in the polar direction, providing excellent convergence to
various black-hole metrics~\cite{Younsi2016}. Since these parametrized
metrics are not the result of a generic parametrized action, no field
equations can be associated to the RZ/KRZ metrics, which thus cannot be
employed in scenarios or simulations where the spacetime is dynamical.

Although the RZ/KRZ parametrizations have been successful in describing
arbitrary black-hole metric in a theory-independent manner, they are
constructed only for the exterior portion of the spacetime, that is, for
the part of the spacetime between the event horizon and spatial infinity
in the case of vacuum spacetimes, or for the part of the spacetime
between the surface of the compact star and spatial infinity in the case
of non-vacuum spacetimes. The interior region, despite not being
necessarily undefined, is disconnected from the exterior by virtue of the
coordinate singularity at the horizon. Numerical simulations, on the
other hand, traditionally make use of ``horizon-penetrating'' (HP)
coordinates, that is, coordinate systems that are regular across the
horizon, so that the inner boundary of the computational domain can be
placed in a region that is causally disconnected from the exterior and
hence does not require special or sophisticated boundary conditions. In
the case of Kerr spacetimes, the most commonly simulated spacetimes for
rotating black holes~\cite{Porth2019_etal}, this is accomplished by
expressing the Kerr metric not in Boyer-Lindquist coordinates (that are
singular at the horizon), but in terms of Kerr-Schild coordinates, hence
obtaining a coordinate mapping that is regular everywhere in the interior
with the obvious exception of the ring singularity. Clearly, it would be
useful to have RZ/KRZ parametrizations that have similar features, that
is, that are regular at the event horizon and in the interior.

A first attempt to derive a version of the KRZ parametrization that is
regular on the event horizon was proposed by Konoplya, Kunz and
Zhidenko~\cite{Konoplya2021}. Although HP in principle, the KRZ
formulation in Ref.~\cite{Konoplya2021} is not useful in practice. This
is because they used a coordinate transformation that alters the
curvature invariants, leading to a Kerr reduction that is not Ricci-flat.
The metric form in Ref.~\cite{Konoplya2021} also has a zero $g_{r r}$
component and hence a zero determinant of the three-metric, incompatible
with a $3+1$ split of spacetime normally employed in numerical simulation
codes. We here present a different HP formulation of the RZ/KRZ
parametrizations with invariant Ricci scalar and nonzero three-metric
determinant. As such, they can be used in numerical simulations
modelling, for instance, the accretion flows onto arbitrary black holes
in alternative theories of gravity. In such scenarios, in fact, the
gravitational mass of the accreting material is many orders of magnitude
smaller than that of the black hole, and the spacetime is therefore
determined by the black hole to a very good approximation\footnote{Taking
for example the case of the accretion flows onto the supermassive black
holes M87*~\cite{Akiyama2019_L1_etal} or Sgr
A*~\cite{EHT_SgrA_PaperI_etal}, the mass of a typical accretion disc with
the external edge at $10^4$ gravitational radii is only $10^{-12}$
($10^{-6}$) that of the central black hole
M87*~\cite{Akiyama2019_L5_etal} (Sgr A*~\cite{EHT_SgrA_PaperV_etal}).
Similar small ratios can be computed in the case of accretion onto X-ray
binaries~\cite{Frank2002}.}. Hence, given a specific physical scenario
where the background spacetime is not influenced by the dynamics of
matter or fields to be evolved (baryonic matter, electromagnetic fields,
radiation fields, scalar fields, etc.), the use of our HP KRZ metric
allows one to simulate black-hole accretion processes by evolving the
corresponding conservation equations of energy, momentum and rest-mass
from large distances down to the black-hole interior without encountering
any coordinate singularity.

The structure of the paper is as follows. In Sec.~\ref{sec:KRZ metric} we
review the basic aspects of the KRZ parametrization for general rotating
black holes. Section~\ref{sec:HP coordinates} reports instead the main
steps needed for the derivation of our HP versions of the KRZ
parametrization in spherical coordinates. We also offer an alternative
Cartesian formulation of the HP KRZ metric in Sec.~\ref{sec:Cartesian
HP}. Sec.~\ref{sec:known BH} applies our HP coordinates to a couple of
examples of rotating black holes. Sec.~\ref{sec:HP RZ} provides a
reduction of the HP KRZ metric to nonrotating black holes. Finally, a
brief discussion of our results is presented in
Sec.~\ref{sec:conclusion}. Hereafter, we adopt a set of units in which
$c = 1 = G$, with $c$ and $G$ being the speed of light and the
gravitational constant, respectively. Furthermore, as usual, we adopt
Greek letters for indices running from $0$ to $3$ and Latin letters for
indices running from $1$ to $3$.

\section{General stationary black-hole metrics}
\label{sec:KRZ metric}

In our construction of HP coordinates of parametrized black-hole metrics,
it is more convenient to start from the case of stationary black holes
and reduce the resulting expressions to the nonrotating case. Hence, we
start from the standard KRZ parametrization~\cite{Konoplya2016a}, where
the spacetime around a general rotating black hole in an arbitrary metric
theory of gravity is stationary and axisymmetric, described by a metric
in the form
\begin{widetext}
\begin{equation}
 \label{eq:KRZ_metric}
 ds^2 = -\frac{N^2 - W^2 \sin^2{\theta}}{K^2} d t^2 - 2 W r
 \sin^2{\theta} d t d \phi + \Sigma \left(\frac{B^2}{N^2} d r^2 + r^2 d
 \theta^2\right) + K^2 r^2 \sin^2{\theta} d \phi^2 \,,
\end{equation}
\end{widetext}
where the coordinates $t$ and $\phi$ are associated with the timelike and
spacelike (azimuthal) Killing vectors marking the spacetime symmetry,
while $r$ and $\theta$ are along the radial and angular directions
perpendicular to $\phi$, so that $(t, r, \theta, \phi)$ form a set of
coordinates that are spherical asymptotically\footnote{In the case of the
Kerr metric, such coordinates are represented by the Boyer-Lindquist
coordinates.}. In the metric~\eqref{eq:KRZ_metric}, the functions $B =
B(r, \theta)$, $N = N(r, \theta)$, $K = K(r, \theta)$, $W = W(r,
\theta)$, and $\Sigma = \Sigma(r, \theta)$ are functions of $r$ and
$\theta$ only, with the latter being defined as
\begin{equation}
 \label{eq:Sigma}
 \Sigma := 1 + \frac{a_*^2 \cos^2{\theta}}{r^2} \,,
\end{equation}
with $a_* := J / M$ being the rotation parameter of the black hole, while
$M$ and $J$ the black-hole mass and angular momentum, respectively.

As a result, the determinant of the KRZ metric \eqref{eq:KRZ_metric} is
then given by
\begin{equation}
 \label{eq:KRZ_det}
 g := \det(g_{\mu \nu}) = -\Sigma^2 B^2 r^4 \sin^2{\theta} \,,
\end{equation}
while the inverse metric can be found by using the identity
$\delta^{\mu}_{\nu} = g^{\mu \lambda} g_{\lambda \nu}$, with the nonzero
components being
\begin{align}
 \label{eq:KRZ_inv_1}
 g^{t t} &= -\frac{g_{\phi \phi}}{g_{t \phi}^2 - g_{t t} g_{\phi \phi}} =
 -\frac{K^2}{N^2} \,, \\
 \label{eq:KRZ_inv_2}
 g^{t \phi} &= \frac{g_{t \phi}}{g_{t \phi}^2 - g_{t t} g_{\phi \phi}} =
 -\frac{W}{N^2 r} \,, \\
 \label{eq:KRZ_inv_3}
 g^{r r} &= \frac{1}{g_{r r}} = \frac{N^2}{\Sigma B^2} \,, \\
 \label{eq:KRZ_inv_4}
 g^{\theta \theta} &= \frac{1}{g_{\theta \theta}} = \frac{1}{\Sigma r^2}
 \,, \\
 \label{eq:KRZ_inv_5}
 g^{\phi \phi} &= -\frac{g_{t t}}{g_{t \phi}^2 - g_{t t} g_{\phi \phi}} =
 \frac{N^2 - W^2 \sin^2{\theta}}{N^2 K^2 r^2 \sin^2{\theta}} \,.
\end{align}

The event horizon of the black hole in the KRZ metric is located at a
surface where the metric component $g_{r r}$ diverges (or $g^{r r}$
vanishes), that is,
\begin{equation}
 \label{eq:horizon}
 N^2(r, \theta) = 0 \,.
\end{equation}
In particular, $r_0$ is the horizon radius in the equatorial plane (\ie
at $\theta = \pi/2$), so that $N^2(r_0, \pi/2) = 0$. The other important
surface of a stationary black hole is represented by the ``static
limit'', and is defined as the surface at which the metric component
$g_{t t}$ vanishes
\begin{equation}
 \label{eq:static_limit}
 N^2 = W^2 \sin^2{\theta} \,.
\end{equation}
The region between these two surfaces is the ergosphere and hence it is
characterized by the condition
\begin{equation}
 \label{eq:ergosphere}
 0 < N^2 < W^2 \sin^2{\theta} \,,
\end{equation}
and, within the Kerr metric, it represents the region where energy and
angular momentum can be extracted from the black hole via the Penrose
process (see, \eg Ref.~\cite{Lasota2014} for a very comprehensive
overview).

The geometry of such axisymmetric spacetime depends on five quantities:
one constant parameter $a_*$, and four functions $B$, $N$, $K$, $W$, that
can be parametrized in terms of expansions in the $r$ and $\theta$
directions. An essential aspect of the RZ, and therefore of the KRZ
framework, is the introduction of a compactified radial coordinate
\begin{equation}
 \label{eq:x_coord}
 \tilde{x} := 1 - \frac{r_0}{r} \,, 
\end{equation}
which maps the black hole exterior $r \in [r_0, \infty)$ to $\tilde{x}
\in [0, 1)$ and hence allows one to impose rather trivially the
asymptotic properties of the spacetime. Similarly, one can introduce the
new angular coordinate
\begin{equation}
 \label{eq:y_coord}
 \tilde{y} := \cos{\theta} \,,
\end{equation}
which maps the polar angle $\theta \in [0, \pi/2]$ to $\tilde{y} \in [1,
0]$. Adopting these new variables, the metric functions in terms of
$(\tilde{x}, \tilde{y})$ of the KRZ metric are expressed after a variable
separation in a product of functions of $\tilde{x}$ and a Taylor series
of $\tilde{y}$
\begin{align}
 \label{eq:y_expan_1}
 N^2(\tilde{x}, \tilde{y}) &= \tilde{x} A_0(\tilde{x}) +
 \sum_{i=1}^{\infty} A_i(\tilde{x}) \tilde{y}^i \,, \\
 \label{eq:y_expan_2}
 B(\tilde{x}, \tilde{y}) &= 1 + \sum_{i=0}^{\infty} B_i(\tilde{x})
 \tilde{y}^i \,, \\
 \label{eq:y_expan_3}
 W(\tilde{x}, \tilde{y}) &= \frac{1}{\Sigma} \sum_{i=0}^{\infty}
 W_i(\tilde{x}) \tilde{y}^i \,, \\
 \label{eq:y_expan_4}
 K^2(\tilde{x}, \tilde{y}) &= 1 + \frac{a_* W}{r} + \frac{1}{\Sigma}
 \sum_{i=0}^{\infty} K_i(\tilde{x}) \tilde{y}^i \,,
\end{align}
and then the functions of only $\tilde{x}$ are expressed as
\begin{widetext}
\begin{align}
 \label{eq:x_expan_1}
 B_i(\tilde{x}) &:= b_{i0} (1 - \tilde{x}) + \tilde{B}_i(\tilde{x}) (1 -
 \tilde{x})^2 \,, \\
 \label{eq:x_expan_2}
 W_i(\tilde{x}) &:= w_{i0} (1 - \tilde{x})^2 + \tilde{W}_i(\tilde{x}) (1
 - \tilde{x})^3 \,, \\
 \label{eq:x_expan_3}
 K_i(\tilde{x}) &:= k_{i0} (1 - \tilde{x})^2 + \tilde{K}_i(\tilde{x}) (1
 - \tilde{x})^3 \,, \\
 \label{eq:x_expan_4}
 A_0(\tilde{x}) &:= 1 - \epsilon_0 (1 - \tilde{x}) + (a_{00} + k_{00} -
 \epsilon_0) (1 - \tilde{x})^2 + \tilde{A}_0(\tilde{x}) (1 - \tilde{x})^3
 \,, \\
 \label{eq:x_expan_4+}
 A_{i \geq 1}(\tilde{x}) &:= K_i(\tilde{x}) + \epsilon_i (1 -
 \tilde{x})^2 + a_{i0} (1 - \tilde{x})^3 + \tilde{A}_i(\tilde{x}) (1 -
 \tilde{x})^4 \,,
\end{align}
\end{widetext}
where the tilded functions $\tilde{A}_i, \tilde{B}_i, \tilde{K}_i$ and
$\tilde{W}_i$ are expressed as Pad{\'e} series in terms of continued
fractions of $\tilde{x}$
\begin{align}
 \label{eq:Pade_expan_1}
 \tilde{A}_i(\tilde{x}) &:= \cfrac{a_{i1}}{1 + \cfrac{a_{i2} \tilde{x}}{1
 + \cfrac{a_{i3} \tilde{x}}{1 + \cdots}}} \,, \\
 \label{eq:Pade_expan_2}
 \tilde{B}_i(\tilde{x}) &:= \cfrac{b_{i1}}{1 + \cfrac{b_{i2} \tilde{x}}{1
 + \cfrac{b_{i3} \tilde{x}}{1 + \cdots}}} \,, \\
 \label{eq:Pade_expan_3}
 \tilde{K}_i(\tilde{x}) &:= \cfrac{k_{i1}}{1 + \cfrac{k_{i2} \tilde{x}}{1
 + \cfrac{k_{i3} \tilde{x}}{1 + \cdots}}} \,, \\
 \label{eq:Pade_expan_4}
 \tilde{W}_i(\tilde{x}) &:= \cfrac{w_{i1}}{1 + \cfrac{w_{i2} \tilde{x}}{1
 + \cfrac{w_{i3} \tilde{x}}{1 + \cdots}}} \,.
\end{align}

Thanks to the superior convergence properties offered by continued
fractions, the expanded metric can approximate deviations from a
Schwarzschild spacetime with the same mass or a Kerr spacetime with the
same spin in terms of a few parameters only. A detailed discussion of the
properties of the expansion in a variety of spacetimes can be found in a
number of recent works~\cite{Nampalliwar2019, Kocherlakota2020,
Kocherlakota2021, Suvorov2021, Abdikamalov2021, Yu2021, Shashank2021,
Kocherlakota2022}.

\section{Horizon-penetrating coordinates}
\label{sec:HP coordinates}

Since the $r r$ component of the metric~\eqref{eq:KRZ_metric} diverges at
the event horizon, we need to find a coordinate system that is HP,
namely, where this function is regular at the event horizon and hence
allows to smoothly join the interior with the exterior.

\subsection{A first Ansatz}
\label{sec:KKZ formulation}

As mentioned in the Introduction, as a first attempt to derive a version
of the KRZ parametrization that is regular on the event horizon,
Konoplya, Kunz and Zhidenko~\cite{Konoplya2021} introduced a new time and
azimuthal variable in an Eddington-Finkelstein-like form
\begin{align}
 \label{eq:coord_transf_1}
 d \hat{t} &= d t + C(r, \theta) d r \,, \\
 \label{eq:coord_transf_2}
 d \hat{\phi} &= d \phi + D(r, \theta) d r \,,
\end{align}
where $C$ and $D$ are smooth functions of $r$ and $\theta$. The new
coordinates $\hat{t}$ and $\hat{\phi}$ defined have now a dependence on
the coordinates $r$ and $\theta$, so that the KRZ
metric~\eqref{eq:KRZ_metric} in these coordinates becomes
\begin{widetext}
\begin{equation}
 \label{eq:KKZ_transf}
 ds^2 = -\frac{N^2 - W^2 \sin^2{\theta}}{K^2} d \hat{t}^2 - 2 W r
 \sin^2{\theta} d \hat{t} d \hat{\phi} + K^2 r^2 \sin^2{\theta} d
 \hat{\phi}^2 + \Sigma r^2 d \theta^2 + g_{r r} d r^2 + 2 g_{r \hat{t}} d
 r d \hat{t} + 2 g_{r \hat{\phi}} d r d \hat{\phi} \,,
\end{equation}
where
\begin{align}
 \label{eq:new_comp_1}
 g_{r r} &= \left(\frac{\Sigma B^2}{N^2} - \frac{N^2}{K^2} C^2\right) +
 \left(\frac{W}{K^2 r} C - D\right)^2 K^2 r^2 \sin^2{\theta} \,, \\
 \label{eq:new_comp_2}
 g_{r \hat{t}} &= \frac{N^2}{K^2} C - \left(\frac{W}{K^2 r} C - D\right)
 W r \sin^2{\theta} \,, \\
 \label{eq:new_comp_3}
 g_{r \hat{\phi}} &= \left(\frac{W}{K^2 r} C - D\right) K^2 r^2
 \sin^2{\theta} \,.
\end{align}
\end{widetext}

Since the functions $C$ and $D$ are arbitrary, they minimised the number
of new terms by imposing a relation
\begin{equation}
 \label{eq:CD_relation}
 D = \frac{W}{K^2 r} C \,.
\end{equation}
As a result, all the associated brackets in
Eqs.~\eqref{eq:new_comp_1}--\eqref{eq:new_comp_3} vanish. To specify the
transformation function $C$, those authors proposed
\begin{equation}
 \label{eq:KKZ_C}
 C = \frac{\sqrt{\Sigma B^2 K^2}}{N^2} \,,
\end{equation}
so that the two terms in $g_{r r}$ cancel out exactly
\begin{equation}
 \label{eq:KKZ_grr}
 g_{r r} = \frac{\Sigma B^2}{N^2} - \frac{N^2}{K^2} C^2 = 0 \,,
\end{equation}
and thus the only new metric component is
\begin{equation}
 \label{eq:KKZ_grt}
 g_{r \hat{t}} = \frac{N^2}{K^2} C = \sqrt{\frac{\Sigma B^2}{K^2}} \,.
\end{equation}

In summary, adopting this new set of variables, and dropping the hat
symbol for the $\hat{t}$ and $\hat{\phi}$ coordinates, the KRZ metric
\eqref{eq:KRZ_metric} is transformed into
\begin{widetext}
\begin{equation}
 \label{eq:KKZ_metric}
 ds^2 = -\frac{N^2 - W^2 \sin^2{\theta}}{K^2} d t^2 + 2
 \sqrt{\frac{\Sigma B^2}{K^2}} d t d r - 2 W r \sin^2{\theta} d t d \phi
 + \Sigma r^2 d \theta^2 + K^2 r^2 \sin^2{\theta} d \phi^2 \,.
\end{equation}
\end{widetext}
Despite taking a simple form, with five nonzero components as in the
original Boyer-Lindquist-like coordinates and is regular at the horizon,
the KRZ metric in these coordinates~\eqref{eq:KKZ_metric} is problematic.
It is easy to realize this by considering the
metric~\eqref{eq:KKZ_metric} when reduced to the case of a Kerr
spacetime. We recall that the Kerr metric in Boyer-Lindquist coordinates
takes the form
\begin{widetext}
\begin{equation}
 \label{eq:Kerr_BL}
 ds^2 = -\left(1 - \frac{2 M r}{\Sigma_\mathrm{K}}\right) d t^2 - \frac{4
 M r}{\Sigma_\mathrm{K}} a_* \sin^2{\theta} d t d \phi +
 \frac{\Sigma_\mathrm{K}}{\Delta} d r^2 + \Sigma_\mathrm{K} d \theta^2 +
 \frac{A}{\Sigma_\mathrm{K}} \sin^2{\theta} d \phi^2 \,,
\end{equation}
\end{widetext}
where the functions
\begin{align}
 \label{eq:Kerr_Sigma}
 \Sigma_\mathrm{K} &:= r^2 + a_*^2 \cos^2{\theta} \,, \\
 \label{eq:Kerr_Delta}
 \Delta &:= r^2 - 2 M r + a_*^2 \,, \\
 \label{eq:Kerr_A}
 A &:= (r^2 + a_*^2)^2 - a_*^2 \Delta \sin^2{\theta} \,,
\end{align}
and where it should be noted that the function $\Sigma_\mathrm{K}$
differs by a quadratic term in the radial coordinate from the
corresponding KRZ function, \ie $\Sigma_\mathrm{K} = \Sigma r^2$
($\Sigma$ is dimensionless while $\Sigma_\mathrm{K}$ has the dimensions
of a squared length).

A direct comparison between Kerr metric~\eqref{eq:Kerr_BL} and the
original KRZ metric presented in Eq.~\eqref{eq:KRZ_metric} implies that,
the free metric functions in the metric \eqref{eq:KKZ_metric} for the
case of a Kerr spacetime are given by
\begin{align}
 \label{eq:KRZ_to_Kerr_1}
 B^2 &= 1 \,, \\
 \label{eq:KRZ_to_Kerr_2}
 N^2 &= \frac{\Delta}{r^2} \,, \\
 \label{eq:KRZ_to_Kerr_3}
 K^2 &= \frac{A}{\Sigma_\mathrm{K} r^2} \,, \\
 \label{eq:KRZ_to_Kerr_4}
 W &= \frac{2 M a_*}{\Sigma_\mathrm{K}} \,.
\end{align}
Substituting the 
relations~\eqref{eq:KRZ_to_Kerr_1}--\eqref{eq:KRZ_to_Kerr_4} in the
transformed HP metric~\eqref{eq:KKZ_metric} yields the KRZ metric reduced
to a Kerr spacetime in the HP coordinates
\begin{widetext}
\begin{equation}
 \label{eq:Kerr_KKZ}
 ds^2 = -\left(1 - \frac{2 M r}{\Sigma_\mathrm{K}}\right) d t^2 + \frac{2
 \Sigma_\mathrm{K}}{\sqrt{A}} d t d r - \frac{4 M r}{\Sigma_\mathrm{K}}
 a_* \sin^2{\theta} d t d \phi + \Sigma_\mathrm{K} d \theta^2 + \frac{A}
 {\Sigma_\mathrm{K}} \sin^2{\theta} d \phi^2 \,.
\end{equation}
\end{widetext}
The Ricci scalar of such a metric does not vanish everywhere, but is
instead given by
\begin{equation}
 \label{eq:KKZ_Ricci}
 R := g^{\mu \nu} R_{\mu \nu} = \frac{a_*^4 \Delta \sin^2{\theta}
 \cos^2{\theta}}{\Sigma_\mathrm{K}^3 A^2} F_1(r, \theta) \,,
\end{equation}
where the function $F_1$ has the radial dependence $F_1(r, \theta)
\propto r^n$ with $n \leqslant 6$. In other words, the metric is flat
only at the horizon, the polar axis and the equatorial plane. Because a
well-posed coordinate transformation cannot change the Ricci-flatness
property of the Kerr solution, this is an indication that the initial
coordinate
transformations~\eqref{eq:coord_transf_1}--\eqref{eq:coord_transf_2} are
not adequate and alternative approaches need to be found.

Besides leading to a Kerr reduction that is not Ricci-flat, the metric in
Eq.~\eqref{eq:KKZ_metric} also suffers from an additional drawback when
it needs to be implemented in numerical simulations (see
Refs.~\cite{Mizuno2018, Kocherlakota2023} for some examples of numerical
simulations of accretion onto black holes in different theories of
gravity). This is because numerical codes solving the equations of
general-relativistic hydrodynamics or magnetohydrodynamics systematically
adopt a $3+1$ split of spacetime where the metric and its inverse have
components given by (see, \eg \cite{MTW1973, Rezzolla_book:2013})
\begin{align}
 \label{eq:3+1_def_1}
 g_{\mu \nu} &= \begin{pmatrix} -\alpha^2 + \beta^k \beta_k & \beta_j \\
 & \\ \beta_i & \gamma_{i j} \end{pmatrix} \,, \\
 \nonumber \\
 \label{eq:3+1_def_2}
 g^{\mu \nu} &= \begin{pmatrix} -1 / \alpha^2 & \beta^j / \alpha^2 \\ &
 \\ \beta^i / \alpha^2 & \gamma^{i j} - \beta^i \beta^j / \alpha^2
 \end{pmatrix} \,,
\end{align}
where $\alpha$ is the (scalar) lapse function, $\boldsymbol{\beta}$ is
the shift vector, and $\boldsymbol{\gamma}$ is the spatial three-metric.
In the $3+1$ split, the determinant of the four-metric $g := \det(g_{\mu
\nu})$ and that of the three-metric $\gamma := \det(\gamma_{i j})$ are
related by the simple expression
\begin{equation}
 \label{eq:det_relation}
 \sqrt{-g} = \alpha \sqrt{\gamma} \,.
\end{equation}
Since the metric in Eq.~\eqref{eq:KKZ_metric} has $g_{r r} = \gamma_{r r}
= 0$, the determinant of the three-metric is zero ($\gamma = 0$), but
that of the four-metric is nonzero as in Eq.~\eqref{eq:KRZ_det}, thus
leaving the lapse function divergent. Overall, therefore, the coordinate
transformation in
Eqs.~\eqref{eq:coord_transf_1}--\eqref{eq:coord_transf_2} leads to a KRZ
form~\eqref{eq:KKZ_metric} that is not useful in practice. In view of
these drawbacks, in the following section we present a KRZ form in HP
coordinates that is regular at the horizon and that can be implemented in
numerical simulations. [See Appendix~\ref{app:MR transformation} for a
discussion about the coordinate transformation leading to the KRZ
formulation in Eq.~\eqref{eq:KKZ_metric}.]

\subsection{A new Ansatz for a subclass of KRZ metric}
\label{sec:separable KRZ}

As anticipated in the previous Section, a different approach to find a
form of the KRZ metric that is HP and leads to a Ricci-flat Kerr
reduction is to start from a formulation of the KRZ expansion where the
metric functions that are themselves separable.  Fortunately, this
problem has already been solved by Konoplya, Stuchl{\'i}k and
Zhidenko~\cite{Konoplya2018}, who have derived a subclass of the KRZ
metric that allows for separation of variables. In such a subclass, the
KRZ functions can be written as
\bigskip
\begin{widetext}
\begin{align}
 \label{eq:KRZ_sub_B}
 B(r, \theta) &=: R_B(r) \,, \\
 \label{eq:KRZ_sub_N}
 N^2(r, \theta) &=: 1 - \frac{R_M(r)}{r} + \frac{a_*^2}{r^2} \,, \\
 \label{eq:KRZ_sub_W}
 W(r, \theta) &= \frac{1}{\Sigma(r, \theta)} \frac{a_* R_M(r)}{r^2} \,,
 \\
 \label{eq:KRZ_sub_K}
 K^2(r, \theta) &= \frac{1}{\Sigma(r, \theta)} \left(1 + \frac{a_*^2}
 {r^2} + \frac{a_*^2 \cos^2{\theta}}{r^2} N^2(r)\right) + \frac{a_* W(r,
 \theta)}{r} \nonumber \\ &= \frac{1}{\Sigma(r, \theta)} \left[1 +
 \frac{a_*^2}{r^2} + \frac{a_*^2}{r^2} \frac{R_M(r)}{r} + \frac{a_*^2
 \cos^2{\theta}}{r^2} \left(1 - \frac{R_M(r)}{r} +
 \frac{a_*^2}{r^2}\right)\right] \nonumber \\ &= \Sigma(r, \theta) +
 \left(1 + \frac{R_M(r)}{\Sigma(r, \theta) r}\right) \frac{a_*^2
 \sin^2{\theta}}{r^2} \,,
\end{align}
\end{widetext}
where Eq.~\eqref{eq:KRZ_sub_N} can be taken as the implicit definition of
the function $R_M(r)$. Note that in this way, four free KRZ functions:
$B$ and $N$ depend on $r$ alone, while the functions $W$ and $K$ have
also a $\theta$-dependence, but only in terms of $\Sigma(r, \theta)$ and
$\sin^2{\theta}$.

We can now proceed with a coordinate transformation that will guarantee
regularity across the horizon by requiring that the transformation
functions $\hat{C}$ and $\hat{D}$ are independent of $\theta$, \ie
$\hat{C} = \hat{C}(r)$, $\hat{D} = \hat{D}(r)$. The differential form is
then reduced to
\begin{align}
 \label{eq:transf_sub_1}
 d \hat{t} &= d t + C(r) d r \,, \\
 \label{eq:transf_sub_2}
 d \hat{\phi} &= d \phi + D(r) d r \,,
\end{align}
so that the transformed metric remains as in
Eqs.~\eqref{eq:KKZ_transf}--\eqref{eq:new_comp_3}. Note that when
choosing the transformation functions $C(r)$ and $D(r)$ to regularize the
metric components in Eqs.~\eqref{eq:new_comp_1}--\eqref{eq:new_comp_3},
we can no longer force the relation~\eqref{eq:CD_relation}, because the
factor $W / (K^2 r)$ is dependent on $\theta$. Thus, none of the brackets
in Eqs.~\eqref{eq:new_comp_1}--\eqref{eq:new_comp_3} vanish, but each of
them has to be regular at the horizon, \ie they cannot contain a factor
$N^2$ in the denominator. To this scope we set
\begin{align}
 \label{eq:HP_C}
 C(r) &= \frac{R_B R_M}{N^2 r} \,, \\
 \label{eq:HP_D}
 D(r) &= \frac{R_B a_*}{N^2 r^2} \,,
 \end{align}
which results in the condition
\begin{align}
 \label{eq:HP_CD_bracket}
 \frac{W}{K^2 r} C - D &= \frac{R_B a_*}{N^2 r^2} \left(\frac{1}{\Sigma
 K^2} \frac{R_M^2}{r^2} - 1\right) \nonumber \\ &= -\frac{R_B a_*}{K^2
 r^2} \left(1 + \frac{R_M}{\Sigma r}\right) \,.
\end{align}

The corresponding modified metric components are then simplified as
\begin{widetext}
\begin{align}
 \label{eq:HP_grr}
 g_{r r} &= \frac{R_B^2}{N^2 K^2} \left(\Sigma K^2 -
 \frac{R_M^2}{r^2}\right) + \frac{R_B^2 a_*^2}{K^2 r^2} \left(1 +
 \frac{R_M}{\Sigma r}\right)^2 \sin^2{\theta} \nonumber \\ &=
 \frac{\Sigma R_B^2}{K^2} \left(1 + \frac{R_M}{\Sigma r}\right) +
 \frac{R_B^2}{K^2} \left(1 + \frac{R_M}{\Sigma r}\right)^2 \frac{a_*^2
 \sin^2{\theta}}{r^2} = \left(1 + \frac{R_M}{\Sigma r}\right) R_B^2 \,,
 \\
 \label{eq:HP_grt}
 g_{r \hat{t}} &= \frac{R_B R_M}{K^2 r} + \frac{R_B R_M}{K^2 r} \left(1 +
 \frac{R_M}{\Sigma r}\right) \frac{a_*^2 \sin^2{\theta}}{\Sigma r^2} =
 \frac{R_B R_M}{\Sigma r} \,, \\
 \label{eq:HP_grphi}
 g_{r \hat{\phi}} &= -\left(1 + \frac{R_M}{\Sigma r}\right) R_B a_*
 \sin^2{\theta} \,,
\end{align}
\end{widetext}
leading to the following form of the separable KRZ metric in HP
coordinates (where we drop the hat on the $t$ and $\phi$ coordinates)
\begin{widetext}
\begin{align}
 \label{eq:HP_KRZ_metric}
 ds^2 = &-\left(1 - \frac{R_M}{\Sigma r}\right) d t^2 + 2 \frac{R_M}
 {\Sigma r} R_B d t d r - 2 \frac{R_M}{\Sigma r} a_* \sin^2{\theta} d t d
 \phi \nonumber \\ &+ \left(1 + \frac{R_M}{\Sigma r}\right) R_B^2 d r^2 -
 2 \left(1 + \frac{R_M}{\Sigma r}\right) R_B a_* \sin^2{\theta} d r d
 \phi + \Sigma r^2 d \theta^2 + K^2 r^2 \sin^2{\theta} d \phi^2 \,.
\end{align}
\end{widetext}

We note that, as already remarked in Ref.~\cite{Konoplya2018}, it is in
principle possible to introduce a new radial coordinate defined through
the differential $d \hat{r} = R_B d r$, such that the function $R_B(r)$
is absorbed into the differential and hence the function $B(\hat{r}) =
1$. While this choice does produce an apparent simplification of the
metric form~\eqref{eq:HP_KRZ_metric}, it does at the cost of introducing
more complex expressions for the radial functions $\Sigma$, $N^2$ and
$K^2$, which have to adopt a definition different from those given in
Eqs.~\eqref{eq:Sigma}, \eqref{eq:KRZ_sub_N} and \eqref{eq:KRZ_sub_K},
and hence containing an additional radial function $R_{\Sigma}(\hat{r})$.
In practice, therefore, the new radial coordinate $\hat{r}$ does not lead
to a significant (or effective) simplification to the
metric~\eqref{eq:HP_KRZ_metric}; for this reason, we will not consider it
further in the following sections.

To calculate the inverse of the metric in Eq.~\eqref{eq:HP_KRZ_metric},
it is useful to first use the following identity
\begin{equation}
 \label{eq:inv_denom}
 g_{t t} g_{r \phi}^2 - g_{t r} g_{t \phi} g_{r \phi} + g_{t r}^2 g_{\phi
 \phi} - g_{t t} g_{r r} g_{\phi \phi} = \Sigma R_B^2 r^2 \sin^2{\theta}
 \,.
\end{equation}
which, in turn, simplifies the expressions of the inverse metric, whose
nonzero components are
\begin{widetext}
\begin{align}
 \label{eq:HP_inv_1}
 g^{t t} &= \frac{g_{r \phi}^2 - g_{r r} g_{\phi \phi}}{g_{t t} g_{r
 \phi}^2 - g_{t r} g_{t \phi} g_{r \phi} + g_{t r}^2 g_{\phi \phi} - g_{t
 t} g_{r r} g_{\phi \phi}} = -\left(1 + \frac{R_M}{\Sigma r}\right) \,,
 \\
 \label{eq:HP_inv_2}
 g^{t r} &= \frac{g_{t r} g_{\phi \phi} - g_{t \phi} g_{r \phi}}{g_{t t}
 g_{r \phi}^2 - g_{t r} g_{t \phi} g_{r \phi} + g_{t r}^2 g_{\phi \phi} -
 g_{t t} g_{r r} g_{\phi \phi}} = \frac{R_M}{\Sigma R_B r} \,, \\
 \label{eq:HP_inv_3}
 g^{r r} &= \frac{g_{t \phi}^2 - g_{t t} g_{\phi \phi}}{g_{t t} g_{r
 \phi}^2 - g_{t r} g_{t \phi} g_{r \phi} + g_{t r}^2 g_{\phi \phi} - g_{t
 t} g_{r r} g_{\phi \phi}} = \frac{N^2}{\Sigma R_B^2} \,, \\
 \label{eq:HP_inv_4}
 g^{r \phi} &= \frac{g_{t t} g_{r \phi} - g_{t r} g_{t \phi}}{g_{t t}
 g_{r \phi}^2 - g_{t r} g_{t \phi} g_{r \phi} + g_{t r}^2 g_{\phi \phi} -
 g_{t t} g_{r r} g_{\phi \phi}} = \frac{a_*}{\Sigma R_B r^2} \,, \\
 \label{eq:HP_inv_5}
 g^{\phi \phi} &= \frac{g_{t r}^2 - g_{t t} g_{r r}}{g_{t t} g_{r \phi}^2
 - g_{t r} g_{t \phi} g_{r \phi} + g_{t r}^2 g_{\phi \phi} - g_{t t} g_{r
 r} g_{\phi \phi}} = \frac{1}{\Sigma r^2 \sin^2{\theta}} \,, \\
 \label{eq:HP_inv_6}
 g^{\theta \theta} &= \frac{1}{g_{\theta \theta}} = \frac{1}{\Sigma r^2}
 \,.
\end{align}
\end{widetext}

Note that none of the metric components in Eqs.~\eqref{eq:HP_KRZ_metric}
and~\eqref{eq:HP_inv_1}--\eqref{eq:HP_inv_6} contains the singular term
$1 / N^2$, thus manifesting the HP nature of such a form of the separable
KRZ metric.

With the HP KRZ metric~\eqref{eq:HP_KRZ_metric}, it is important to check
whether it yields a Ricci-flat Kerr reduction. We manage to obtain an
expression for the Ricci scalar while keeping $R_B(r)$ and $R_M(r)$
unspecified
\begin{widetext}
\begin{equation}
 \label{eq:HP_KRZ_Ricci}
 R := g^{\mu \nu} R_{\mu \nu} = \frac{1}{\Sigma^3 R_B^3 r^6} \left[R_B
 (R_B^2 - 1) F_2(r, \theta) + R_B (2 R_M' + R_M'' r) F_3(r, \theta) +
 R_B' r F_4(r, \theta)\right] \,,
\end{equation}
\end{widetext}
where the radial dependence in the functions $F_i$ with $i = 2, 3, 4$ is
given by $F_i(r, \theta) \propto r^n$ with $n \leqslant 4$, and a prime
indicates a radial derivative. Clearly, under the conditions that $R_B =
1$ and $R_M = \mathrm{const}$ (see Sec.~\ref{sec:known BH}), the Ricci
scalar vanishes, as expected for the Kerr metric. Moreover,
expression~\eqref{eq:HP_KRZ_Ricci} does not change if evaluated in the
original coordinates, \ie the standard KRZ metric~\eqref{eq:KRZ_metric}
with separability constraints in
Eqs.~\eqref{eq:KRZ_sub_B}--\eqref{eq:KRZ_sub_K}, thus proving that the
coordinate transformation chosen in
Eqs.~\eqref{eq:transf_sub_1}--\eqref{eq:transf_sub_2} has the desired
properties of providing an HP form of the KRZ metric that is Ricci flat
when reduced to a Kerr black hole.

To facilitate a direct implementation in numerical codes, we report here
explicitly the $3+1$ metric components of \eqref{eq:HP_KRZ_metric}
and~\eqref{eq:HP_inv_1}--\eqref{eq:HP_inv_6}
\begin{align}
 \label{eq:HP_3+1_1}
 \gamma_{r r} &= \left(1 + \frac{R_M}{\Sigma r}\right) R_B^2 \,, \\
 \label{eq:HP_3+1_2}
 \gamma_{r \phi} &= -\left(1 + \frac{R_M}{\Sigma r}\right) R_B a_*
 \sin^2{\theta} \,, \\
 \label{eq:HP_3+1_3}
 \gamma_{\theta \theta} &= \Sigma r^2 \,, \\
 \label{eq:HP_3+1_4}
 \gamma_{\phi \phi} &= K^2 r^2 \sin^2{\theta} \,, \\
 \label{eq:HP_3+1_5}
 \beta_{r} &= \frac{R_B R_M}{\Sigma r} \,, \\
 \label{eq:HP_3+1_6}
 \beta_{\phi} &= -\frac{R_M}{\Sigma r} a_* \sin^2{\theta} \,, \\
 \label{eq:HP_3+1_7}
 \alpha^2 &= \frac{1}{1 + R_M / (\Sigma r)} \,,
\end{align}
while the corresponding inverse is
\begin{align}
 \label{eq:HP_3+1_inv_1}
 \gamma^{r r} &= \frac{K^2}{\Sigma R_B^2} \frac{1}{1 + R_M / (\Sigma r)}
 \,, \\
 \label{eq:HP_3+1_inv_2}
 \gamma^{r \phi} &= \frac{a_*}{\Sigma R_B r^2} \,, \\
 \label{eq:HP_3+1_inv_3}
 \gamma^{\theta \theta} &= \frac{1}{\Sigma r^2} \,, \\
 \label{eq:HP_3+1_inv_4}
 \gamma^{\phi \phi} &= \frac{1}{\Sigma r^2 \sin^2{\theta}} \,, \\
 \label{eq:HP_3+1_inv_5}
 \beta^{r} &= \frac{R_M}{\Sigma R_B r} \frac{1}{1 + R_M / (\Sigma r)} \,,
 \\
 \label{eq:HP_3+1_inv_6}
 \beta^{\phi} &= 0 \,, \\
 \label{eq:HP_3+1_inv_7}
 \beta^k \beta_k &= \frac{R_M^2 / (\Sigma^2 r^2)}{1 + R_M / (\Sigma r)}
 \,.
\end{align}

\section{Cartesian form of the KRZ metric in HP coordinates}
\label{sec:Cartesian HP}

It is not unusual that general-relativistic codes implement black-hole
metrics in Cartesian coordinates as these, by construction, remove
possible coordinate singularities at the polar axis and are generally
easier to handle in fully numerical-relativity codes (see,
\eg~\cite{Porth2019_etal} for a comparison of different codes). Hence, it
is convenient to derive expressions of the KRZ metric in HP coordinates
also in Cartesian coordinates and to this scope, we take inspiration by
the mathematical path followed when expressing the Kerr metric in
Cartesian Kerr-Schild coordinates. More specifically, we start with a
``Kerr-Schild decomposition'' where the metric is split into two parts
\begin{equation}
 \label{eq:KS_decomp}
 g_{\mu \nu} = \mathring{g}_{\mu \nu} + f \ell_{\mu} \ell_{\nu} \,,
\end{equation}
where $\mathring{g}_{\mu \nu}$ is a reference metric with a particularly
simple form, $f$ is a scalar function, and $\ell_{\mu}$ is a null vector
with respect to the full metric, \ie
\begin{equation}
 \label{eq:null_vec}
 g^{\mu \nu} \ell_{\mu} \ell_{\nu} = 0 \,.
\end{equation}
The decomposition~\eqref{eq:KS_decomp} is totally generic and hence
employable in any coordinate system. However, it is easier for us to try
a Kerr-Schild form starting from spherical coordinates and then transform
over to Cartesian ones. Inspired by what done for the Kerr solution in
Kerr-Schild coordinates, we fix the scalar function $f$ in
Eq.~\eqref{eq:KS_decomp} to be
\begin{equation}
 \label{eq:f_assum}
 f := \frac{R_M}{\Sigma r} \,.
\end{equation}
From the relevant metric functions
\begin{align}
 \label{eq:gr_form_1}
 g_{r r} &= \left(1 + \frac{R_M}{\Sigma r}\right) R_B^2 \,, \\
 \label{eq:gr_form_2}
 g_{r \theta} &= 0 \,, \\
 \label{eq:gr_form_3}
 g_{r \phi} &= -\left(1 + \frac{R_M}{\Sigma r}\right) R_B a_*
 \sin^2{\theta} \,,
\end{align}
the null vector is assumed to have components
\begin{equation}
 \label{eq:null_assum}
 \ell_{r} = R_B \,, \qquad \ell_{\theta} = 0 \,, \qquad
 \ell_{\phi} = -a_* \sin^2{\theta} \,.
\end{equation}
As a result, we can solve Eq.~\eqref{eq:null_vec} to obtain the component
$\ell_{t}$, namely, 
\begin{widetext}
\begin{align}
 \label{eq:solve_lt}
 g^{\mu \nu} \ell_{\mu} \ell_{\nu} &= g^{t t} \ell_{t} \ell_{t} + 2 g^{t
 r} \ell_{t} \ell_{r} + g^{r r} \ell_{r} \ell_{r} + 2 g^{r \phi} \ell_{r}
 \ell_{\phi} + g^{\phi \phi} \ell_{\phi} \ell_{\phi} \nonumber \\ &=
 -\left(1 + \frac{R_M}{\Sigma r}\right) \ell_{t} \ell_{t} + \frac{2 R_M}
 {\Sigma r} \ell_{t} + \left(1 - \frac{R_M}{\Sigma r}\right) = 0 \,,
\end{align}
\end{widetext}
with the solutions being
\begin{equation}
 \label{eq:lt}
 \ell_{t} = 1 \,, \qquad \mathrm{and} \qquad
 \ell_{t} = -\frac{\Sigma r - R_M}{\Sigma r + R_M} \,.
\end{equation}
For simplicity, hereafter we will consider the simpler solution $\ell_{t}
= 1$. The remaining part of the metric has then the following form
\begin{align}
 \label{eq:g0_metric_1}
 \mathring{g}_{t t} &= -1 \,, \\
 \label{eq:g0_metric_2}
 \mathring{g}_{t r} &= 0 \,, \\
 \label{eq:g0_metric_3}
 \mathring{g}_{t \phi} &= 0 \,, \\
 \label{eq:g0_metric_4}
 \mathring{g}_{r r} &= R_B^2 \,, \\
 \label{eq:g0_metric_5}
 \mathring{g}_{r \phi} &= -R_B a_* \sin^2{\theta} \,, \\
 \label{eq:g0_metric_6}
 \mathring{g}_{\theta \theta} &= \Sigma r^2 \,, \\
 \label{eq:g0_metric_7}
 \mathring{g}_{\phi \phi} &= (r^2 + a_*^2) \sin^2{\theta} \,.
\end{align}

Next, when wishing to express the Kerr metric in Cartesian Kerr-Schild
coordinates we can adopt a set of coordinates defined as
\begin{align}
 \label{eq:Cart_x}
 x &:= \sqrt{r^2 + a_*^2} \sin{\theta} \cos{\left(\phi +
 \arctan{\frac{a_*}{r}}\right)} \,, \\
 \label{eq:Cart_y}
 y &:= \sqrt{r^2 + a_*^2} \sin{\theta} \sin{\left(\phi +
 \arctan{\frac{a_*}{r}}\right)} \,, \\
 \label{eq:Cart_z}
 z &:= r \cos{\theta} \,.
\end{align}
Employing the Kerr-Schild decomposition~\eqref{eq:KS_decomp} and the
Cartesian coordinates~\eqref{eq:Cart_x}--\eqref{eq:Cart_z}, we can obtain
the separable KRZ metric in Cartesian HP coordinates. In particular, the
corresponding null vector $\ell_{\mu}$ will have spatial components
\begin{align}
 \label{eq:null_Cart_1}
 \ell_{x} &= \frac{x}{\Sigma r} \left[R_B - \frac{a_*^2 (x^2 + y^2)}{(r^2
 + a_*^2)^2}\right] + \frac{a_* y}{r^2 + a_*^2} \,, \\
 \label{eq:null_Cart_2}
 \ell_{y} &= \frac{y}{\Sigma r} \left[R_B - \frac{a_*^2 (x^2 + y^2)}{(r^2
 + a_*^2)^2}\right] - \frac{a_* x}{r^2 + a_*^2} \,, \\
 \label{eq:null_Cart_3}
 \ell_{z} &= \frac{z}{\Sigma r} \left[\frac{r^2 + a_*^2}{r^2} R_B -
 \frac{a_*^2 (x^2 + y^2)}{r^2 (r^2 + a_*^2)}\right] \,.
\end{align}
where $\Sigma$ is now expressed as
\begin{equation}
 \label{eq:Sigma_Cart}
 \Sigma = 1 + \frac{a_*^2 z^2}{r^4} \,,
\end{equation}
and $r$ is implicitly determined by the relation
\begin{equation}
 \label{eq:xyz-r}
 \frac{x^2 + y^2}{r^2 + a_*^2} + \frac{z^2}{r^2} = 1 \,,
\end{equation}
which marks the ring singularity when $z = 0$.

Similarly, the components of the reference metric $\mathring{g}_{\mu
\nu}$ will be given by
\begin{widetext}
\begin{align}
 \label{eq:g0_Cart_1}
 \mathring{g}_{t x} &= \mathring{g}_{t y} = \mathring{g}_{t z} = 0 \,, \\
 \label{eq:g0_Cart_2}
 \mathring{g}_{x x} &= \frac{x^2}{\Sigma^2 r^2} \left[R_B^2 - \frac{a_*^2
 (x^2 + y^2)}{(r^2 + a_*^2)^2} (2 R_B - 1) + \Sigma \frac{z^2 (r^2 +
 a_*^2)}{r^2 (x^2 + y^2)}\right] + \frac{y^2}{x^2 + y^2} - \frac{a_*}
 {\Sigma r} \frac{2 x y}{r^2 + a_*^2} (R_B - 1) \,, \\
 \label{eq:g0_Cart_3}
 \mathring{g}_{y y} &= \frac{y^2}{\Sigma^2 r^2} \left[R_B^2 - \frac{a_*^2
 (x^2 + y^2)}{(r^2 + a_*^2)^2} (2 R_B - 1) + \Sigma \frac{z^2 (r^2 +
 a_*^2)}{r^2 (x^2 + y^2)}\right] + \frac{x^2}{x^2 + y^2} - \frac{a_*}
 {\Sigma r} \frac{2 x y}{r^2 + a_*^2} (R_B - 1) \,, \\
 \label{eq:g0_Cart_4}
 \mathring{g}_{z z} &= \frac{z^2}{\Sigma^2 r^2} \left[\left(\frac{r^2 +
 a_*^2}{r^2}\right)^2 R_B^2 - \frac{a_*^2 (x^2 + y^2)}{r^4} (2 R_B - 1) +
 \Sigma \frac{r^2 (x^2 + y^2)}{z^2 (r^2 + a_*^2)}\right] \,, \\
 \label{eq:g0_Cart_5}
 \mathring{g}_{x y} &= \frac{x y}{\Sigma^2 r^2} \left[R_B^2 - \frac{a_*^2
 (x^2 + y^2)}{(r^2 + a_*^2)^2} (2 R_B - 1) + \Sigma \frac{z^2 (r^2 +
 a_*^2)}{r^2 (x^2 + y^2)}\right] - \frac{x y}{x^2 + y^2} - \frac{a_*}
 {\Sigma r} \frac{x^2 - y^2}{r^2 + a_*^2} (R_B - 1) \,, \\
 \label{eq:g0_Cart_6}
 \mathring{g}_{x z} &= \frac{x z}{\Sigma^2 r^2} \left[\frac{r^2 + a_*^2}
 {r^2} R_B^2 - \frac{a_*^2 (x^2 + y^2)}{r^2 (r^2 + a_*^2)} (2 R_B - 1) -
 \Sigma\right] + \frac{a_*}{\Sigma r} \frac{y z}{r^2} (R_B - 1) \,, \\
 \label{eq:g0_Cart_7}
 \mathring{g}_{y z} &= \frac{y z}{\Sigma^2 r^2} \left[\frac{r^2 + a_*^2}
 {r^2} R_B^2 - \frac{a_*^2 (x^2 + y^2)}{r^2 (r^2 + a_*^2)} (2 R_B - 1) -
 \Sigma\right] - \frac{a_*}{\Sigma r} \frac{x z}{r^2} (R_B - 1) \,.
\end{align}
\end{widetext}
It is then not difficult to verify that the null
condition~\eqref{eq:null_vec} still holds in terms of the full
metric~\eqref{eq:KS_decomp}. In particular the case of $R_B(r) = 1$, the
null vector is reduced to
\begin{align}
 \label{eq:null_Cart_B1_1}
 \ell_{x} &= \frac{r x + a_* y}{r^2 + a_*^2} \,, \\
 \label{eq:null_Cart_B1_2}
 \ell_{y} &= \frac{r y - a_* x}{r^2 + a_*^2} \,, \\
 \label{eq:null_Cart_B1_3}
 \ell_{z} &= \frac{z}{r} \,,
\end{align}
and the reference metric can be dramatically simplified since
$\mathring{g}_{\mu \nu} = \eta_{\mu \nu}$, the Minkowski metric for flat
spacetime. For compactness, we will not present here the $3+1$
expressions for the KRZ metric in Cartesian HP coordinates, which we
however report in Appendix~\ref{app:Cartesian HP 3+1}.

\section{Reduction to known stationary black holes}
\label{sec:known BH}

We next show how various known metrics describing stationary black holes
can be represented within the framework of KRZ metric in HP coordinates
by suitably choosing specific expressions for the metric functions
$R_B(r)$ and $R_M(r)$.

\subsection{Kerr metric}
\label{sec:Kerr}

Starting from the generic metric~\eqref{eq:HP_KRZ_metric}, the reduction
to the Kerr spacetime is obtained when fixing
\begin{align}
 \label{eq:RB_Kerr}
 R_B(r) &= 1 \,, \\
 \label{eq:RM_Kerr}
 R_M(r) &= 2 M \,,
\end{align}
so that Eqs.~\eqref{eq:KRZ_sub_B}--\eqref{eq:KRZ_sub_K} reduce to
Eqs.~\eqref{eq:KRZ_to_Kerr_1}--\eqref{eq:KRZ_to_Kerr_4}. Under the
conditions~\eqref{eq:RB_Kerr}--\eqref{eq:RM_Kerr}, the HP KRZ
metric~\eqref{eq:HP_KRZ_metric} becomes
\begin{widetext}
\begin{align}
 \label{eq:HP_Kerr}
 ds^2 = &-\left(1 - \frac{2 M r}{\Sigma_\mathrm{K}}\right) d t^2 +
 \frac{4 M r}{\Sigma_\mathrm{K}} d t d r - \frac{4 M r}
 {\Sigma_\mathrm{K}} a_* \sin^2{\theta} d t d \phi \nonumber \\ &+
 \left(1 + \frac{2 M r}{\Sigma_\mathrm{K}}\right) d r^2 - 2 \left(1 +
 \frac{2 M r}{\Sigma_\mathrm{K}}\right) a_* \sin^2{\theta} d r d \phi +
 \Sigma_\mathrm{K} d \theta^2 + \frac{A}{\Sigma_\mathrm{K}}
 \sin^2{\theta} d \phi^2 \,,
\end{align}
\end{widetext}
which is precisely the Kerr-Schild form of Kerr metric and hence
Ricci-flat. The fact that the reduction of the KRZ metric in HP
coordinates~\eqref{eq:HP_KRZ_metric} leads to the Kerr-Schild metric in
the case of the Kerr solution represents a very important feature, as it
considerably simplifies the comparison between Kerr black holes and
other, non-Kerr but stationary black holes described by the KRZ
metric~\eqref{eq:HP_KRZ_metric}.

The Cartesian HP form of the metric satisfies the relation
\begin{equation}
 \label{eq:KS_decomp_Kerr}
 g_{\mu \nu} = \eta_{\mu \nu} + \frac{2 M r}{\Sigma_\mathrm{K}}
 \ell_{\mu} \ell_{\nu} \,,
\end{equation}
with the null four-vector having components
\begin{equation}
 \label{eq:null_vec_Kerr}
 \ell_{\mu} = \left(1, \frac{r x + a_* y}{r^2 + a_*^2}, \frac{r y - a_*
 x}{r^2 + a_*^2}, \frac{z}{r}\right) \,.
\end{equation}

\subsection{Kerr-Newman metric}
\label{sec:Kerr-Newman}

Similarly, the Kerr-Newman metric in HP coordinates can be obtained from
the generic HP KRZ metric~\eqref{eq:HP_KRZ_metric} after setting
\begin{align}
 \label{eq:RB_KN}
 R_B(r) &= 1 \,, \\
 \label{eq:RM_KN}
 R_M(r) &= 2 M - \frac{Q^2}{r} \,,
\end{align}
where $Q$ is the electric charge of the black hole. The KRZ functions are
then modified as
\begin{align}
 \label{eq:KRZ_to_KN_1}
 B^2 &= 1 \,, \\
 \label{eq:KRZ_to_KN_2}
 N^2 &= \frac{\Delta}{r^2} + \frac{Q^2}{r^2} \,, \\
 \label{eq:KRZ_to_KN_3}
 K^2 &= \frac{A}{\Sigma_\mathrm{K} r^2} - \frac{Q^2 a_*^2}
 {\Sigma_\mathrm{K} r^4} \,, \\
 \label{eq:KRZ_to_KN_4}
 W &= \frac{2 M a_*}{\Sigma_\mathrm{K}} - \frac{Q^2 a_*}
 {\Sigma_\mathrm{K} r} \,.
\end{align}
As a result, the reduction of the HP KRZ metric~\eqref{eq:HP_KRZ_metric}
in the case of the Kerr-Newman solution is
\begin{widetext}
\begin{align}
 \label{eq:HP_KN}
 ds^2 = &-\left(1 - \frac{2 M r - Q^2}{\Sigma_\mathrm{K}}\right) d t^2 +
 2 \frac{2 M r - Q^2}{\Sigma_\mathrm{K}} d t d r - 2 \frac{2 M r - Q^2}
 {\Sigma_\mathrm{K}} a_* \sin^2{\theta} d t d \phi \nonumber \\ &+
 \left(1 + \frac{2 M r - Q^2}{\Sigma_\mathrm{K}}\right) d r^2 - 2 \left(1
 + \frac{2 M r - Q^2}{\Sigma_\mathrm{K}}\right) a_* \sin^2{\theta} d r d
 \phi + \Sigma_\mathrm{K} d \theta^2 + \frac{A r^2 - Q^2 a_*^2}
 {\Sigma_\mathrm{K} r^2} \sin^2{\theta} d \phi^2 \,,
\end{align}
\end{widetext}
which, to the best of our knowledge, has not been presented before in the
literature (an HP formulation of the Kerr-Newmann solution in Cartesian
coordinates can be found in Ref.~\cite{Debney1969}, while a version in
null coordinates has been presented in Ref.~\cite{Adamo2014}). The
corresponding Cartesian HP form is then
\begin{equation}
 \label{eq:KS_decomp_KN}
 g_{\mu \nu} = \eta_{\mu \nu} + \frac{2 M r - Q^2}{\Sigma_\mathrm{K}}
 \ell_{\mu} \ell_{\nu} \,,
\end{equation}
with the null vector~\eqref{eq:null_vec_Kerr} unchanged.

\subsection{Rotating dilaton metric}
\label{sec:dilaton}

For a rotating dilaton black hole characterized by rotation parameter
$a_*$ and dilaton parameter $b_*$, the KRZ functions have the following
form
\begin{align}
 \label{eq:KRZ_to_dilat_1}
 B^2 &= \frac{r^2}{r^2 + b_*^2} \,, \\
 \label{eq:KRZ_to_dilat_2}
 N^2 &= 1 - \frac{2 M \rho}{r^2} + \frac{a_*^2}{r^2} =:
 \frac{\Delta_\mathrm{d}}{r^2} \,, \\
 \label{eq:KRZ_to_dilat_3}
 K^2 &= \frac{\Sigma_\mathrm{K}}{r^2} + \left(1 + \frac{2 M \rho}
 {\Sigma_\mathrm{K}}\right) \frac{a_*^2 \sin^2{\theta}}{r^2} =:
 \frac{A_\mathrm{d}}{\Sigma_\mathrm{K} r^2} \,, \\
 \label{eq:KRZ_to_dilat_4}
 W &= \frac{2 M \rho a_*}{\Sigma_\mathrm{K} r} \,,
\end{align}
which imply
\begin{align}
 \label{eq:RB_dilat}
 R_B(r) &= \sqrt{\frac{r^2}{r^2 + b_*^2}} \,, \\
 \label{eq:RM_dilat}
 R_M(r) &= \frac{2 M \rho}{r} \,,
\end{align}
where
\begin{equation}
 \label{eq:rho_dilat}
 \rho := \sqrt{r^2 + b_*^2} - b_* \,.
\end{equation}
Therefore, the corresponding HP KRZ metric~\eqref{eq:HP_KRZ_metric}
describing a rotating dilaton black hole is
\begin{widetext}
\begin{align}
 \label{eq:HP_dilat}
 ds^2 = &-\left(1 - \frac{2 M \rho}{\Sigma_\mathrm{K}}\right) d t^2 +
 \frac{4 M \rho}{\Sigma_\mathrm{K}} \sqrt{\frac{r^2}{r^2 + b_*^2}} d t d
 r - \frac{4 M \rho}{\Sigma_\mathrm{K}} a_* \sin^2{\theta} d t d \phi
 \nonumber \\ &+ \left(1 + \frac{2 M \rho}{\Sigma_\mathrm{K}}\right)
 \frac{r^2}{r^2 + b_*^2} d r^2 - 2 \left(1 + \frac{2 M \rho}
 {\Sigma_\mathrm{K}}\right) \sqrt{\frac{r^2}{r^2 + b_*^2}} a_*
 \sin^2{\theta} d r d \phi + \Sigma_\mathrm{K} d \theta^2 +
 \frac{A_\mathrm{d}}{\Sigma_\mathrm{K}} \sin^2{\theta} d \phi^2 \,.
\end{align}
\end{widetext}
The Cartesian HP form again satisfies the decomposition
\begin{equation}
 \label{eq:KS_decomp_dilat}
 g_{\mu \nu} = \mathring{g}_{\mu \nu} + \frac{2 M \rho}
 {\Sigma_\mathrm{K}} \ell_{\mu} \ell_{\nu} \,,
\end{equation}
with the null vector and reference metric as expressed in
Eqs.~\eqref{eq:null_Cart_1}--\eqref{eq:null_Cart_3} and
Eqs.~\eqref{eq:g0_Cart_1}--\eqref{eq:g0_Cart_7}.

\section{Reduction to static black holes}
\label{sec:HP RZ}

The KRZ framework describing generic stationary black
holes~\cite{Konoplya2016a} inherited much of the mathematical properties
that make it so efficient from the previous RZ approach describing
generic static black holes~\cite{Rezzolla2014}. We complete our treatment
of the HP formulation of the KRZ metric by considering also the simpler,
but often more transparent, case of nonrotating spacetimes. To this
scope, we recall that the RZ metric, to which the KRZ metric reduces in
the case of spherically symmetric spacetimes, takes the
form~\cite{Rezzolla2014}
\begin{equation}
 \label{eq:RZ_metric}
 ds^2 = -N^2 \, d t^2 + \frac{B^2}{N^2} d r^2 + r^2 d \theta^2 + r^2
 \sin^2{\theta} d \phi^2 \,.
\end{equation}
A rapid comparison with the KRZ metric~\eqref{eq:KRZ_metric} indicates
that the two are equivalent if
\begin{align}
 \label{eq:KRZ_to_RZ_1}
 a_* & = 0 \,, \\
 \label{eq:KRZ_to_RZ_2}
 \Sigma & = 1 \,, \\
 \label{eq:KRZ_to_RZ_3}
 W & = 0 \,, \\
 \label{eq:KRZ_to_RZ_4}
 K^2 & = 1 \,,
\end{align}
and thus the remaining metric functions become
\begin{align}
 \label{eq:RZ_B}
 B(r) &=: R_B(r) \,, \\
 \label{eq:RZ_N}
 N^2(r) &=: 1 - \frac{R_M(r)}{r} \,,
\end{align}
so that the HP KRZ metric~\eqref{eq:HP_KRZ_metric} reduces to the RZ
metric in HP coordinates
\begin{widetext}
\begin{align}
 \label{eq:HP_RZ_metric}
 ds^2 &= -\left(1 - \frac{R_M}{r}\right) d t^2 + \frac{2 R_B R_M}{r} d t
 d r + \left(1 + \frac{R_M}{r}\right) R_B^2 d r^2 + r^2 d \theta^2 + r^2
 \sin^2{\theta} d \phi^2 \nonumber \\ &= -N^2 d t^2 + 2 \left(1 -
 N^2\right) B d t d r + \left(2 - N^2\right) B^2 d r^2 + r^2 d \theta^2 +
 r^2 \sin^2{\theta} d \phi^2 \,.
\end{align}
\end{widetext}
Note that the HP RZ metric~\eqref{eq:HP_RZ_metric} has nontrivial inverse
metric components given by
\begin{align}
 \label{eq:HP_RZ_inv_1}
 g^{t t} &= -(2 - N^2) \,, \\
 \label{eq:HP_RZ_inv_2}
 g^{t r} &= \frac{1 - N^2}{B} \,, \\
 \label{eq:HP_RZ_inv_3}
 g^{r r} &= \frac{N^2}{B^2} \,, \\
 \label{eq:HP_RZ_inv_4}
 g^{\theta \theta} &= \frac{1}{r^2} \,, \\
 \label{eq:HP_RZ_inv_5}
 g^{\phi \phi} &= \frac{1}{r^2 \sin^2{\theta}} \,.
\end{align}
It is interesting to note that the metric~\eqref{eq:HP_RZ_metric}
effectively represents the generalisation to arbitrary static spacetimes
of the well-known Eddington-Finkelstein coordinates employed for a
Schwarzschild black hole~\cite{Eddington1923, Finkelstein58, MTW1973}. In
the case of a Schwarzschild black hole, in fact, the RZ metric functions
are fixed to be
\begin{align}
 \label{eq:RZ_to_Schw_1}
 B^2 & = 1 \,, \\
 \label{eq:RZ_to_Schw_2}
 N^2 & = 1 - \frac{2M}{r} \,,
\end{align}
so that the metric~\eqref{eq:HP_RZ_metric} takes the form
\begin{widetext}
\begin{equation}
 \label{eq:Schw_EF}
 ds^2 = -\left(1 - \frac{2 M}{r}\right) d t^2 + \frac{4 M}{r} d t d r +
 \left(1 + \frac{2 M}{r}\right) d r^2 + r^2 d \theta^2 + r^2
 \sin^2{\theta} d \phi^2 \,,
\end{equation}
\end{widetext}
which indeed corresponds to the Schwarzschild solution in (in-going)
Eddington-Finkelstein coordinates.

Finally, and in analogy with what done with the KRZ metric in HP
coordinates, we provide below explicit expressions for the corresponding
components when the metric is written in a $3+1$ split:
\begin{align}
 \label{eq:HP_RZ_3+1_1}
 \gamma_{i j} &= \mathrm{diag} \left(B^2 (2 - N^2), r^2, r^2
 \sin^2{\theta}\right) \,, \\
 \label{eq:HP_RZ_3+1_2}
 \gamma^{i j} &= \mathrm{diag} \left(\frac{1}{B^2 (2 - N^2)}, \frac{1}
 {r^2}, \frac{1}{r^2 \sin^2{\theta}}\right) \,, \\
 \label{eq:HP_RZ_3+1_3}
 \beta_{r} &= B (1 - N^2) \,, \\
 \label{eq:HP_RZ_3+1_4}
 \beta^{r} &= \frac{1 - N^2}{B (2 - N^2)} \,, \\
 \label{eq:HP_RZ_3+1_5}
 \beta^k \beta_k &= \frac{(1 - N^2)^2}{2 - N^2} \,, \\
 \label{eq:HP_RZ_3+1_6}
 \alpha^2 &= \frac{1}{2 - N^2} \,.
\end{align}

\section{Conclusion}
\label{sec:conclusion}

As new and unprecedented observations of stellar-mass and supermassive
black holes are now becoming available, and as the number of new
alternative theories of gravity is increasing steadily, it is clear that
parametrized approaches to describe the metric of static and stationary
black holes represent a very effective approach to extract agnostically
information from such observations.

In this spirit, we have here considered the family of the
Rezzolla-Zhidenko (RZ) and Konoplya-Rezzolla-Zhidenko (KRZ)
parametrizations of black-hole spacetimes -- which are able to reproduce
arbitrary black-hole spacetimes with percent precision and only a few
coefficients -- as an reference approach to study the phenomenology of
matter near black holes. These parametrizations, however, have
constructed only for the exterior portion of the spacetime that, in the
case of black holes, spans the region between the event horizon and
spatial infinity. For this reason, they are not optimal for actual
numerical simulations as those studying the accretion onto supermassive
black holes, which instead make use of ``horizon-penetrating'' (HP)
coordinates that are well defined in the black-hole interior and up to
the physical singularity.

We have therefore discussed two different Ansaetze for the derivation
of an HP form of the KRZ metric. The first approach leads to a regular
version of the KRZ metric, but suffer from having a nonzero Ricci scalar
when reduced to the Kerr limit. The violation of this constraint is to be
found in the coordinate transformations employed that are too general and
unrestricted to provide at the same time regularity and a Ricci-flat Kerr
reduction. To compensate for these shortcomings, we have considered an
alternative Ansatz which starts from a subclass of the KRZ metric that
already allows for separation of variables of the Hamilton-Jacobi
equations. We then show that in such a subclass, the KRZ metric can be
written in HP form that shares many similarities with the Kerr-Schild
decomposition. In order to facilitate the use of this HP form of the KRZ
metric, we have derived the corresponding expressions in $3+1$
decompositions of the spacetime when employing either spherical or
Cartesian coordinates. Finally, we have shown that when reduced to the
case of Kerr and Schwarzschild black holes, the HP form of the KRZ metric
reduces respectively to the Kerr-Schild representation of the Kerr
spacetime and to the Eddington-Finkelstein formulation of the
Schwarzschild spacetime. This highlights that in the relevant limit, the
HP KRZ formulation leads to Ricci-flat spacetimes.

Because these parametrized metrics will allow to model accretion flows
onto arbitrary black holes in alternative theories of gravity, they have
the potential of being a very effective tool to extract important
information from the astronomical observations of black holes.

\begin{acknowledgments}
It is a pleasure to thank A. Cruz-Osorio, I. Kalpa Dihingia, Y. Mizuno,
and K. Moriyama for useful discussions. We also thank C. Ecker for help
with the calculation of the Ricci scalar and P. Kocherlakota for
insightful discussions on the various coordinate transformations
considered here. Support comes from the ERC Advanced Grant ``JETSET:
Launching, propagation and emission of relativistic jets from binary
mergers and across mass scales'' (Grant No. 884631). LR acknowledges the
Walter Greiner Gesellschaft zur F\"orderung der physikalischen
Grundlagenforschung e.V. through the Carl W. Fueck Laureatus Chair.
\end{acknowledgments}

\appendix

\section{$3+1$ form of the KRZ metric in Cartesian HP coordinates}
\label{app:Cartesian HP 3+1}

Here we report the $3+1$ expressions for the KRZ metric in Cartesian HP
coordinates. According to Sec.~\ref{sec:Cartesian HP} and
Eqs.~\eqref{eq:3+1_def_1}--\eqref{eq:3+1_def_2}, the three-metric in the
Cartesian HP coordinates have components
\begin{widetext}
\begin{align}
 \label{eq:gamma_Cart_1}
 \gamma_{x x} &= \frac{x^2}{\Sigma^2 r^2} \left[R_B^2 - \frac{a_*^2 (x^2
 + y^2)}{(r^2 + a_*^2)^2} (2 R_B - 1) + \Sigma \frac{z^2 (r^2 + a_*^2)}
 {r^2 (x^2 + y^2)}\right] + \frac{y^2}{x^2 + y^2} - \frac{a_*}{\Sigma r}
 \frac{2 x y}{r^2 + a_*^2} (R_B - 1) + \frac{R_M}{\Sigma r} \ell_{x}^2
 \,, \\
 \label{eq:gamma_Cart_2}
 \gamma_{y y} &= \frac{y^2}{\Sigma^2 r^2} \left[R_B^2 - \frac{a_*^2 (x^2
 + y^2)}{(r^2 + a_*^2)^2} (2 R_B - 1) + \Sigma \frac{z^2 (r^2 + a_*^2)}
 {r^2 (x^2 + y^2)}\right] + \frac{x^2}{x^2 + y^2} - \frac{a_*}{\Sigma r}
 \frac{2 x y}{r^2 + a_*^2} (R_B - 1) + \frac{R_M}{\Sigma r} \ell_{y}^2
 \,, \\
 \label{eq:gamma_Cart_3}
 \gamma_{z z} &= \frac{z^2}{\Sigma^2 r^2} \left[\left(\frac{r^2 + a_*^2}
 {r^2}\right)^2 R_B^2 - \frac{a_*^2 (x^2 + y^2)}{r^4} (2 R_B - 1) +
 \Sigma \frac{r^2 (x^2 + y^2)}{z^2 (r^2 + a_*^2)}\right] + \frac{R_M}
 {\Sigma r} \ell_{z}^2 \,, \\
 \label{eq:gamma_Cart_4}
 \gamma_{x y} &= \frac{x y}{\Sigma^2 r^2} \left[R_B^2 - \frac{a_*^2 (x^2
 + y^2)}{(r^2 + a_*^2)^2} (2 R_B - 1) + \Sigma \frac{z^2 (r^2 + a_*^2)}
 {r^2 (x^2 + y^2)}\right] - \frac{x y}{x^2 + y^2} - \frac{a_*}{\Sigma r}
 \frac{x^2 - y^2}{r^2 + a_*^2} (R_B - 1) + \frac{R_M}{\Sigma r} \ell_{x}
 \ell_{y} \,, \\
 \label{eq:gamma_Cart_5}
 \gamma_{x z} &= \frac{x z}{\Sigma^2 r^2} \left[\frac{r^2 + a_*^2}{r^2}
 R_B^2 - \frac{a_*^2 (x^2 + y^2)}{r^2 (r^2 + a_*^2)} (2 R_B - 1) -
 \Sigma\right] + \frac{a_*}{\Sigma r} \frac{y z}{r^2} (R_B - 1) +
 \frac{R_M}{\Sigma r} \ell_{x} \ell_{z} \,, \\
 \label{eq:gamma_Cart_6}
 \gamma_{y z} &= \frac{y z}{\Sigma^2 r^2} \left[\frac{r^2 + a_*^2}{r^2}
 R_B^2 - \frac{a_*^2 (x^2 + y^2)}{r^2 (r^2 + a_*^2)} (2 R_B - 1) -
 \Sigma\right] - \frac{a_*}{\Sigma r} \frac{x z}{r^2} (R_B - 1) +
 \frac{R_M}{\Sigma r} \ell_{y} \ell_{z} \,;
\end{align}
the shift vector $\beta_{i}$ is proportional to $\ell_{i}$, the spatial
part of the null vector
\begin{align}
 \label{eq:beta_Cart_1}
 \beta_{x} = \frac{R_M}{\Sigma r} \ell_{x} &= \frac{R_M x}{\Sigma^2 r^2}
 \left[R_B - \frac{a_*^2 (x^2 + y^2)}{(r^2 + a_*^2)^2}\right] +
 \frac{R_M}{\Sigma r} \frac{a_* y}{r^2 + a_*^2} \,, \\
 \label{eq:beta_Cart_2}
 \beta_{y} = \frac{R_M}{\Sigma r} \ell_{y} &= \frac{R_M y}{\Sigma^2 r^2}
 \left[R_B - \frac{a_*^2 (x^2 + y^2)}{(r^2 + a_*^2)^2}\right] -
 \frac{R_M}{\Sigma r} \frac{a_* x}{r^2 + a_*^2} \,, \\
 \label{eq:beta_Cart_3}
 \beta_{z} = \frac{R_M}{\Sigma r} \ell_{z} &= \frac{R_M z}{\Sigma^2 r^2}
 \left[\frac{r^2 + a_*^2}{r^2} R_B - \frac{a_*^2 (x^2 + y^2)}{r^2 (r^2 +
 a_*^2)}\right] \,;
\end{align}
\end{widetext}
the lapse function remains as
\begin{equation}
 \label{eq:alpha2_Cart}
 \alpha^2 = \frac{1}{1 + R_M / (\Sigma r)} \,.
\end{equation}

\section{A different Ansatz}
\label{app:MR transformation}

For completeness, and setting aside the problematic aspects related to
the $3+1$ decomposition of the KRZ form~\eqref{eq:KKZ_metric}, we here
provide a potential explanation about why it leads to a Kerr reduction
that is not Ricci flat. We believe the origin of this behaviour is to be
found in the incomplete coordinate transformation in
Eqs.~\eqref{eq:coord_transf_1}--\eqref{eq:coord_transf_2}. To see this,
it is sufficient to integrate them and obtain
\begin{align}
 \label{eq:transf_int_1}
 \hat{t} &= t + \hat{C}(r, \theta) \,, \\
 \label{eq:transf_int_2}
 \hat{\phi} &= \phi + \hat{D}(r, \theta) \,,
\end{align}
where
\begin{align}
 \label{eq:transf_func_C^}
 \hat{C}(r, \theta) &:= \int C(r, \theta) d r \,, \\
 \label{eq:transf_func_D^}
 \hat{D}(r, \theta) &:= \int D(r, \theta) d r \,.
\end{align}
Taking again the differential of
Eqs.~\eqref{eq:transf_func_C^}--\eqref{eq:transf_func_D^} we obtain
\begin{align}
 \label{eq:transf_full_1}
 d \hat{t} &= d t + C_{r}(r, \theta) d r + C_{\theta}(r, \theta) d \theta
 \,, \\
 \label{eq:transf_full_2}
 d \hat{\phi} &= d \phi + D_{r}(r, \theta) d r + D_{\theta}(r, \theta) d
 \theta \,,
\end{align}
where
\begin{align}
 \label{eq:Cr_Ctheta}
 C_{r} := \frac{\partial \hat{C}}{\partial r} \,, \qquad &\qquad
 C_{\theta} := \frac{\partial \hat{C}}{\partial \theta} \,, \\
 \label{eq:Dr_Dtheta}
 D_{r} := \frac{\partial \hat{D}}{\partial r} \,, \qquad &\qquad
 D_{\theta} := \frac{\partial \hat{D}}{\partial \theta} \,.
\end{align}

Clearly, Eqs.~\eqref{eq:transf_full_1}--\eqref{eq:transf_full_2} are
different from the original
ansatz~\eqref{eq:coord_transf_1}--\eqref{eq:coord_transf_2} because of
the additional $d \theta$ term, thus indicating that such ansatz is not
general enough. This consideration motivates us to use the full
transformation in Eqs.~\eqref{eq:transf_full_1}--\eqref{eq:transf_full_2}
to obtain a novel form of the KRZ metric in HP coordinates. After some
algebra, the transformed KRZ metric has the form
\begin{widetext}
\begin{align}
 \label{eq:KKZ_full_metric}
 ds^2 = &-\frac{N^2 - W^2 \sin^2{\theta}}{K^2} d \hat{t}^2 - 2 W r
 \sin^2{\theta} d \hat{t} d \hat{\phi} + K^2 r^2 \sin^2{\theta} d
 \hat{\phi}^2 \nonumber \\ &+ g_{r r} d r^2 + 2 g_{r \hat{t}} d r d
 \hat{t} + 2 g_{r \hat{\phi}} d r d \hat{\phi} + 2 g_{r \theta} d r d
 \theta + g_{\theta \theta} d \theta^2 + 2 g_{\theta \hat{t}} d \theta d
 \hat{t} + 2 g_{\theta \hat{\phi}} d \theta d \hat{\phi} \,,
\end{align}
where
\begin{align}
 \label{eq:full_comp_1}
 g_{r r} &= \left(\frac{\Sigma B^2}{N^2} - \frac{N^2}{K^2} C_{r}^2\right)
 + \left(\frac{W}{K^2 r} C_{r} - D_{r}\right)^2 K^2 r^2 \sin^2{\theta}
 \,, \\
 \label{eq:full_comp_2}
 g_{r \hat{t}} &= \frac{N^2}{K^2} C_{r} - \left(\frac{W}{K^2 r} C_{r} -
 D_{r}\right) W r \sin^2{\theta} \,, \\
 \label{eq:full_comp_3}
 g_{r \hat{\phi}} &= \left(\frac{W}{K^2 r} C_{r} - D_{r}\right) K^2 r^2
 \sin^2{\theta} \,, \\
 \label{eq:full_comp_4}
 g_{r \theta} &= -\frac{N^2}{K^2} C_{r} C_{\theta} + \left(\frac{W}{K^2
 r} C_{r} - D_{r}\right) \left(\frac{W}{K^2 r} C_{\theta} -
 D_{\theta}\right) K^2 r^2 \sin^2{\theta} \,, \\
 \label{eq:full_comp_5}
 g_{\theta \theta} &= \left(\Sigma r^2 - \frac{N^2}{K^2}
 C_{\theta}^2\right) + \left(\frac{W}{K^2 r} C_{\theta} -
 D_{\theta}\right)^2 K^2 r^2 \sin^2{\theta} \,, \\
 \label{eq:full_comp_6}
 g_{\theta \hat{t}} &= \frac{N^2}{K^2} C_{\theta} - \left(\frac{W}{K^2 r}
 C_{\theta} - D_{\theta}\right) W r \sin^2{\theta} \,, \\
 \label{eq:full_comp_7}
 g_{\theta \hat{\phi}} &= \left(\frac{W}{K^2 r} C_{\theta} -
 D_{\theta}\right) K^2 r^2 \sin^2{\theta} \,.
\end{align}
Using now the same relation between $C_{r}$ and $D_{r}$ adopted in
Eq.~\eqref{eq:CD_relation}, the transformed metric
components~\eqref{eq:full_comp_1}--\eqref{eq:full_comp_7} become
\begin{align}
 \label{eq:alt_comp_1}
 g_{r r} &= \frac{\Sigma B^2}{N^2} - \frac{N^2}{K^2} C_{r}^2 \,, \\
 \label{eq:alt_comp_2}
 g_{r \hat{t}} &= \frac{N^2}{K^2} C_{r} \,, \\
 \label{eq:alt_comp_3}
 g_{r \hat{\phi}} &= 0 \,, \\
 \label{eq:alt_comp_4}
 g_{r \theta} &= -\frac{N^2}{K^2} C_{r} C_{\theta} \,, \\
 \label{eq:alt_comp_5}
 g_{\theta \theta} &= \left(\Sigma r^2 - \frac{N^2}{K^2}
 C_{\theta}^2\right) + \left(\frac{W}{K^2 r} C_{\theta} -
 D_{\theta}\right)^2 K^2 r^2 \sin^2{\theta} \,, \\
 \label{eq:alt_comp_6}
 g_{\theta \hat{t}} &= \frac{N^2}{K^2} C_{\theta} - \left(\frac{W}{K^2 r}
 C_{\theta} - D_{\theta}\right) W r \sin^2{\theta} \,, \\
 \label{eq:alt_comp_7}
 g_{\theta \hat{\phi}} &= \left(\frac{W}{K^2 r} C_{\theta} -
 D_{\theta}\right) K^2 r^2 \sin^2{\theta} \,.
\end{align}
\end{widetext}

The exact form of the functions $C_{\theta}$ and $D_{\theta}$ is still
unknown, but could be uniquely determined by the choices made for the
functions $C_{r}$ and $D_{r}$ to guarantee the regularity of the metric
at the event horizon. For instance, we could fix the function $C_{r}$,
integrate it in the radial direction to obtain $\hat{C}(r, \theta)$ and
then get $C_{\theta}$ after differentiating in the polar direction.
However, this is possible in practice only for the special
class of functions for which $\hat{C}(r, \theta)$ is separable, \ie for
$\hat{C}(r, \theta) = \Phi(r) \Psi(\theta)$. While it is in principle
possible to proceed in this manner, and this may be explored in future,
the Ansatz made in Sec.~\ref{sec:separable KRZ} to restrict the KRZ to
metric functions that are themselves separable has turned out to be the
simplest and most effective in practice.

\bibliographystyle{apsrev4-2}
%

\end{document}